# Optics-free imaging of complex, non-sparse QR-codes with Deep Neural Networks

EVAN SCULLION,[1] SOREN NELSON,[1] AND RAJESH MENON[1,*]

[1]Department of Electrical & Computer Engineering, University of Utah, Salt Lake City, UT 84112, USA
*rmenon@eng.utah.edu

**Abstract:** We demonstrate optics-free imaging of complex QR-codes using a bare image sensor and a trained artificial neural network (ANN). The ANN is trained to interpret the raw sensor data for human visualization. The image sensor is placed at a specified gap from the QR code. We studied the robustness of our approach by experimentally testing the output of the ANNs with system perturbations of this gap, and the translational and rotational alignments of the QR code to the image sensor. Our demonstration opens us the possibility of using completely optics-free cameras for application-specific imaging of complex, non-sparse objects.

Lensless cameras [1] have been studied extensively because they enable "seeing" around corners either with "natural or accidental" cameras [2] or with active illumination, [3] and generally promise smaller, simpler and cheaper cameras.[4] Phase-only diffractive masks with no optics have been used for 3D imaging, [5] absorption-free color and multi-spectral imaging. [6] Lensless imaging can be applied in microscopy for deep-brain imaging as well. [7,8] By replacing all optics with a transparent window we have shown "see-through" computational cameras. [9] Although there are many other examples of lensless imaging with coherent light, here we are interested in only cameras that perform imaging of incoherent light for generality. Machine-learning has also been widely applied to lensless imaging, primarily for image interpretation for human consumption [10] and less widely for image classification or inferencing directly from the raw (non-human) images [11].

However, cameras that are completely free of any optics have not received significant attention so far. We demonstrated imaging with an optics-free camera (using only the bare image sensor) of simple objects [12]. Machine-learning was used to perform classification of these simple objects without image-reconstructions for human consumption [13]. Such non-anthropomorphic cameras promise enhanced privacy, among other advantages. However, it is not clear, if more complex objects could be imaged in the same manner. Here, we demonstrate an optics-free camera that can reconstruct QR-codes with 29 X 29 pixels, a relatively complex and greatly useful object. Rather than using regularization-based singular-value decomposition, here we utilize a deep artificial-neural network (ANN) to perform the conversion from the Machine ("raw sensor") image to human-readable form. We note that such conversion may be unnecessary in the future, when only machine inferencing is required.

As before, our experiment is performed with a bare image sensor (Mini-2MP-Plus, Arducam) placed at a distance z away from a liquid-crystal display (LCD, Acer G276HL 1920 X 1080). The QR code is displayed on the LCD as illustrated in Fig. 1a. The sizes of the QR code and the sensor are 6mm X 6mm and 6mm X 6mm, respectively. The QR code (29 X 29 boxes) is created using Python ("qrcode" library), with randomly generated 10-character strings.

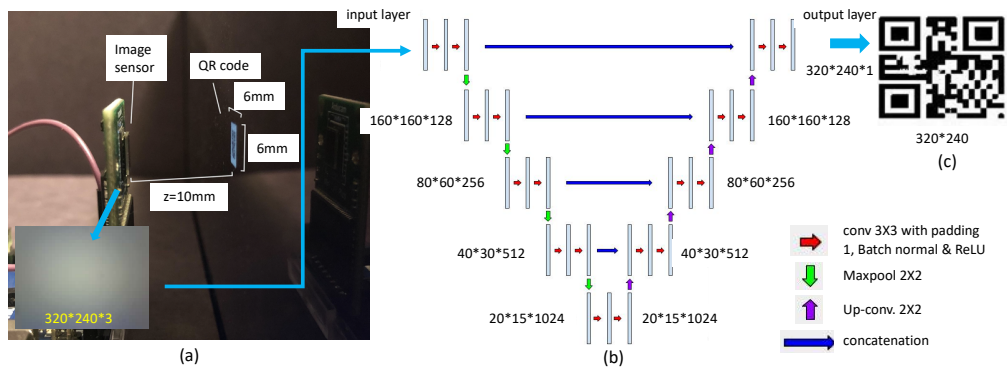

Fig. 1. Details of Experiments. (a) Photograph of setup. (b) Architecture of artificial-neural network. (c) Exemplary output image. The corresponding input image from the sensor is shown in (a).

Previously, we used the singular-value decomposition (SVD) method with regularization to invert a transfer function in order to obtain the images for human consumption from the raw data [12]. Here, we train artificial neural networks (ANNs) to achieve the same results. ANNs have the main advantage that they can scale to larger images and higher resolutions in a more efficient fashion. Furthermore, their performance can be enhanced by additional data and transfer learning. Finally, ANNs could be adapted to perform inferencing directly from the raw data and bypass the image reconstruction step entirely.

We utilized an ANN based on the encoder-decoder architecture as shown in Fig. 1b [10]. One ANN was trained for each value of z=1mm, 5mm and 10mm. Each ANN was comprised of 66 hidden layers and trained on 100 passes over a set size of 20,000 training images. Subsequently, the trained network was validated with a set of 5,000 images, which were excluded from the training set. Each batch size had 10 frames from the sensor. Each frame image was comprised of 3 color channels, each of size 320 X 240 sensor pixels (object pixel size = 0.198mm). The output of the ANN was a single frame of size 320 X 240 pixels. The frame sizes at each major layer of the network is illustrated in Fig. 1b.

The structural similarity index (SSIM) was used as the figure of merit to measure the performance of each ANN. The SSIM is a measure of the fidelity of the reconstructed image with respect to the original (reference). The SSIM was averaged over all the images in each epoch. One ANN for each of z=1mm, 5mm and 10mm was trained. The training and validation averaged SSIM were plotted as functions of epoch in Figs. 2a and 2b, respectively. The smallest value of z performs the best. This is expected, since in any optics-free system free-space propagation over longer distances will tend to increase the mixing of the spatial details of the image. The ANNs at z=1mm, 5mm and 10mm showed best average SSIMs of 87%, 84% and 78%, respectively. Exemplary images reconstructed by each of these ANNs are illustrated in Fig. 2c. Clearly, good quality reconstruction of the QR code is obtained at z=1mm. However, we note that reconstruction is not of sufficient fidelity for a conventional QR-code scanner to identify.

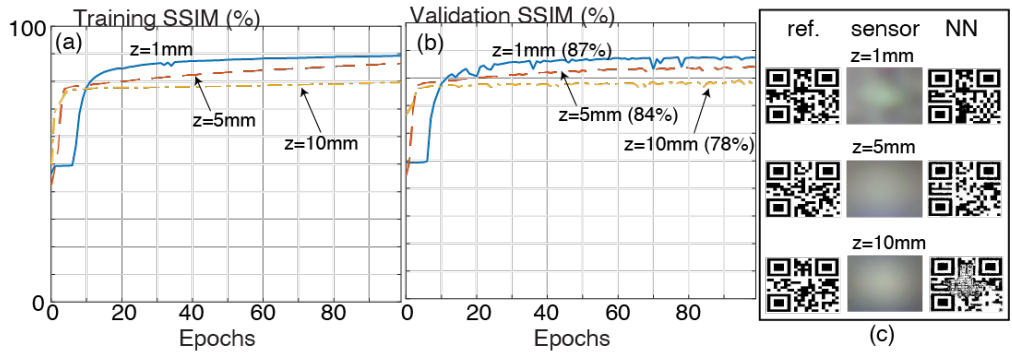

Fig. 2. Results from the ANNs. (a) Training SSIM, (b) validation SSIM (maximum values are shown in parenthesis) and (c) exemplary reference, sensor and output images for the 3 ANNs (z=1mm, 5mm and 10mm).

The sensitivity of our optics-free camera is important for practical applications. In order to study this, we experimentally collected images while varying the gap (z) from 1mm to 10mm (40 images were captured at each gap), and then used each of the previously trained ANNs to reconstruct the images, and computed the average SSIM at each gap (Fig. 3). The changing gap is equivalent to defocus in a lensed camera. As expected, each ANN performs best at the gap that it was trained on. The rate of degradation of SSIM with z seems to be similar for all 3 ANNs. We can conclude from this study that precision of -/+ 0.5mm should be sufficient to maintain SSIM to within ~10% of its peak value.

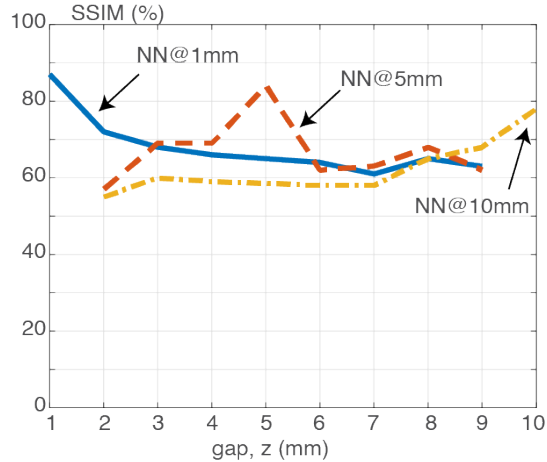

Fig. 3. Robustness of trained ANNs to defocus. As expected, each ANN performs best (highest SSIM) at the gap where it was trained.

Another important parameter to vary is the translation in the XY plane between the QR code relative to the sensor. We studied this by capturing 40 images at each shift corresponding to -/+ 1mm in each direction as illustrated in Fig. 4a. The average SSIM for each ANN is then plotted. The SSIM averaged over all such translated images is also noted above each plot. Our general conclusion is that a shift of less than 1mm is required to maintain SSIM in order to generate usable images of the QR codes. Finally, the impact of rotation of the QR code relative to the sensor along the 3 possible axes was studied as summarized in Fig. 4b. The QR code was rotated along each of the axes by 1 degree and 40 different images were collected for each such rotation. Then, we reconstructed the images using the previously trained ANNs and plotted the average SSIM. The decrease in SSIM is fairly small for all cases, which suggests that the trained ANNs are fairly robust.

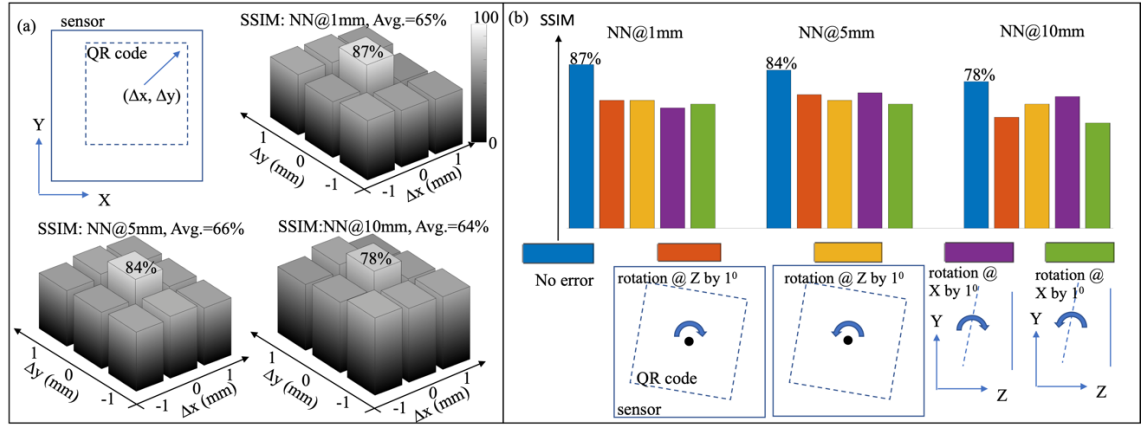

Fig. 4. Robustness of trained ANNs to errors in (a) translation and (b) rotation. The QR code and the sensor are depicted via dashed and solid lines, respectively.

In conclusion, we demonstrated an optics-free camera comprised of a trained artificial neural network and a bare image sensor that is able to convert the raw sensor frames to human-interpretable images of QR codes. Such optics-free cameras have the potential to enable ultra-thin, lightweight, and inexpensive application-specific imaging.

## Funding, acknowledgments, and disclosures

*Funding*

National Science Foundation (NSF) (1533611). University of Utah Undergraduate Research Opportunities Program (UROP).

*Acknowledgments*

We would like to thank Z. Pan and R. Guo for fruitful discussion, and assistance with experiments and software.

*Disclosures*

RM: University of Utah (P).

## References

1. V. Boominathan, J. K. Adams, M. S. Asif, B. W. Avants, J. T. Robinson, R. G. Baraniuk, A. C. Sankaranarayanan and A. Veeraraghavan, "Lensless imaging: A computational renaissance," IEEE Signal Processing Magazine, 33(5) 23-35 (2016).
2. A. Torralba and W. T. Freeman, "Accidental pinhole and pinspeck cameras: revealing the scene outside the picture," IEEE Computer Vision and Pattern Recognition (CVPR) 2012.
3. A. Velten, T. Willwacher, O. Gupta, A. Veeraraghavan, M. G. Bawendi and R. Raskar, "Recovering three-dimensional shpare around a corner using ultrafast time-of-flight imaging," Nat. Comms. 3, 745 (2012).
4. P. R. Gill and D. G. Stork, Lensless ultra-miniature imagers using odd-symmetry spiral phase gratings, Comput. Opt. Sens. Imag., 2013.
5. N. Antipa, G. Kuo, R. Heckel, B. Mildenhall, E. Bostan, R. Ng and L. Waller, "DiffuserCam: lensless single-exposure 3D imaging," Optica 5(1) 1-9 (2018).
6. P. Wang and R. Menon, "Ultra-high sensitivity color imaging via a transparent diffractive-filter array and computational optics," *Optica* 2(11) 933-939 (2015).
7. Ganghun Kim and R. Menon, "An ultra-small 3D computational microscope," *Appl. Phys. Lett.* **105** 061114 (2014).

8. G. Kim, N. Nagarajan, E. Pastuzyn, K. Jenks, M. Capecchi, J. Sheperd and R. Menon,"Deep-brain imaging via epi-fluorescence computational cannula microscopy," *Scientific Reports*, 7:44791 DOI: 10.1038/srep44791 (2016).
9. G. Kim and R. Menon, "Computational imaging enables a "see-through" lensless camera," *Opt. Exp.* 26(18) 22826-22836 (2018)
10. G. Barbastathis, A. Ozcan, and G. Situ, "On the use of deep learning for computational imaging," Optica 6, 921-943 (2019).
11. Z. Pan, B. Rodriguez and R. Menon, "Machine-learning enables Image Reconstruction and Classification in a "see-through" camera," *OSA Continuum* (*in press*).
12. G. Kim, K. Isaacson, R. Palmer and R. Menon, "Lensless photography with only an image sensor," *Appl. Opt.* 56(23),6450-6456 (2017).
13. G. Kim, S. Kapetanovic, R. Palmer and R. Menon, "Lensless-camera based machine learning for image classification," a*rXiv:1709.00408* [cs.CV] (2017).